# Issues for Using Semantic Modeling to Represent Mechanisms

Robert B. Allen
rba@boballen.info

**Abstract:** Mechanisms are a fundamental concept in many areas of science. Nonetheless, there has been little effort to develop structures to represent mechanisms. We explore the issues in developing a basic semantic modeling framework for describing some types of mechanisms. We draw together threads from a number of different approaches and then consider two examples. From this survey, we propose a rich Semantic Modeling Framework (SMF) based on Transitionals and hierarchies of Aggregates and Mechanisms, which could be implemented with the XFO programming environment. Potentially, the framework will be useful for developing direct-representation scientific research reports and community models.

## 1 INTRODUCTION

### 1.1 Mechanisms

Models of mechanisms are a foundation of modern science [8, 9, 16]. They are used in natural sciences ranging from biology and chemistry to geology and meteorology, as well as for cognitive science and social science.

- Mechanisms are entities and activities organized to produce regular changes from start or set-up to finish or termination conditions. ([16] p3)
- Mechanisms are regular in that they always or for the most part work in the same way under the same conditions. The regularity is exhibited in the typical way that the mechanism runs from beginning to end; what makes it regular is the productive continuity between stages. ([16] p3)

Models of mechanisms represent hypotheses that can facilitate understanding of the actions of complex systems. Mechanism models of well-understood phenomena can provide a useful alternative to explanations based purely on statistical regularities. A correlation not based on a plausible mechanism is dubious. In order to infer causal relationships in a complex data set such as in medical diagnosis, it is helpful to have *a priori* candidate mechanisms [17]. Structured mechanism representations can also help to explain the results of machine learning.

Mechanisms are characterized as containing "four basic features: (1) a phenomenon, (2) parts, (3) causings, and (4) organization".[1] These features form a natural foundation for representations;[2] nonetheless, there are challenges in representing mechanisms. For instance, representations are models and, thus, never exactly reproduce reality. Indeed, they often idealize or simplify real-world constraints. Such simplification may even include falsification as Galileo famously did in descriptions of motion that ignored friction.

---

[1] Mechanisms in Science, https://plato.stanford.edu/entries/science-mechanisms/ [Section 2]
[2] [25] identifies several types of models (Galilean, Minimal, and Multiple Models). Because representations of mechanisms are a kind of model, these types may also be applied to mechanism representations.

## 1.2 Rich Semantic Modeling

To represent mechanisms, a framework should support structured coordination of activities and entities. It is helpful to use a controlled vocabulary for the elements of a mechanism that is also used in other applications. Ontologies extend controlled vocabularies by allowing inheritance and specifying relationships among the terms. An upper ontology defines the types of entities that make up reality. We have worked with the Basic Formal Ontology (BFO2) [7], an upper ontology that is widely used in biomedicine[3].

BFO2 is based on realist Aristotelian principles. At the top level, BFO2 distinguishes between Continuants and Occurrents. Continuants are further divided into Independent Continuants and Dependent Continuants. Notable among the Independent Continuants are Material Entities which include Objects, Fiat Object Parts, and Object Aggregates. Notable among the Dependent Continuants are Relational Queries and Functions. In addition to Continuants, there are Occurrents which are Entities which change through time. BFO2 ontologies are "Is_A" hierarchies of terms which map to the upper ontology entities. The Open Biomedical Ontology Foundry[4] (OBO) is a large collection of reference and domain ontologies based on the BFO2 framework.

As an upper ontology, BFO2's goal is to describe the basic units of reality. It does not show how those entities interact. We have proposed extending BFO2 with a model layer [5] which describes the interaction of entities. For instance, we extended BFO2 with components such as Thick Independent Continuants, Transitionals, and control structures (looping, conditionals). We developed a semantic modeling programming environment which we call the Extensible Formal Ontology (XFO) [6]. XFO runs in Python; it implements program statements by changing the links between entities. Furthermore, XFO programs support Object-Oriented Modeling[5]. The strong form includes independent agents which interact in a Microworld via message passing. Thus, a mechanism specification can be considered a program.

## 1.3 Other Approaches

The Function-Behavior-Structure (FBS) approach models the way the FBS factors are coordinated in mechanical processes. In the FBS approach, a Function is roughly comparable to a Mechanism while Behaviors and Structure are related to Causings and Organization. FBS is often used to support system design activities. FBS is notable because it supports the description of the dynamics of mechanical systems, although there are no large, generic FBS controlled vocabularies. The FBS Ontology [12] is focused more on the design process than on applications.

The Systems Biology Markup Language (SBML) is a large, general, and powerful programming environment for developing simulations of biological activities. However, it does not focus on mechanisms per se. The closest work may be by [15], who considered applying the FBS model in systems biology and has many insights, but this approach does not appear to have been widely adopted.

## 1.4 Goals and Roadmap

In this paper, we develop the Semantic Modeling Framework (SMF) for describing mechanisms that can be applied fairly easily and should be able to be semantically validated. A standard structure for describing mechanisms should be helpful for indexing them in a way that can take advantage of the rich semantic modeling provided by systems such as OBO. Our focus here is on relatively straightforward physical mechanisms; thus, we do not consider stochastic mechanisms or social mechanisms.

---

[3] For [16], "Mechanisms are composed of both *entities* (with their properties) and *activities*. Activities are the producers of change. Entities are the things that engage in activities." (Entities in [16] are analogous to Independent Continuants in BFO2.)

[4] http://www.obofoundry.org/

[5] We distinguish between Object-Oriented Design (OOD) and Object-Oriented Analysis (OOA). OOD is widely used and focuses on using object-oriented principles for the design of systems. OOA is more relevant for descriptions of science and history.

In Section 2, we consider Thick Aggregates. In Section 3, we consider Transitionals, while in Section 4 we examine two examples, and in Section 5 we introduce the SMF.

## 2 THICK AGGREGATES (TAs)

An aggregate is a group of often loosely related objects. In BFO2, Object Aggregates are groups of BFO2:Objects.[6] In this paper, we posit that all Independent Continuants are Object Aggregates. Thus, we simplify the distinction that BFO2 makes between Fiat Object Parts, Objects, and Object Aggregates. Complex objects would be hierarchies of Object Aggregates. We assume all Object Aggregates are composed only of other Object Aggregates (e.g., molecules are composed of atoms, atoms are composed of subatomic particles, and so forth).[7]

Among other advantages, this structure is similar to classes in object-oriented languages. To the extent possible, we propose treating all objects defined in the OBO ontologies as Object Aggregates. Further, we define Thick Aggregates (TAs)[8] as Object Aggregates which include interrelated parts and are associated with specific Transitionals. The TA and Transitionals may then be executed as part of a mechanism. In other cases, several independent TAs interact to create a mechanism.

### 2.1 Parts

Complex objects and the mechanisms in which they engage are made up of parts. While in many cases it may seem obvious what the parts are, in other cases it is less clear. It would be helpful to have some criteria for determining the parts. First, parts should be modular units [26]. In addition, functional parts must be able to be manipulated enough to determine whether they have a distinct role in the mechanism. In most cases, parts should be stable and able to be manipulated through interventions. Structural parts may be defined as material entities which maintain the spatial positions allowing the functional parts to interoperate.[9]

The BFO2 foundational relationship "Continuant Part Of" can be used to link the TAs to their parts. The relationships between the parts can be understood by examining the Transitionals in which they participate, and the interaction of the parts can be obtained by examining the entire description of the mechanism. To ensure that the distributed specification is complete and accurate, the representation of the mechanism could be compiled.

In this paper, we focus on material Object Aggregates, which have physical qualities such as mass. For example, because gravity is related to mass, gravitational attraction among several Material Entities could be considered a Relational Quality (RQ). We also allow physical aggregates to have spatial coordinates in a Microworld (see Section 3.4) as a quality; this allows the specification of location, shape, and orientation. There is considerable interest in Object-Oriented Physical Modeling (OOPM) [14], particularly in mechanical engineering with its focus on physical structures and mechanisms. Potentially, rich semantic modeling would complement that work.

### 2.2 States

Because we focus on Transitionals as state changes, we need a clear definition of states. There are several ways for a TA (including parts) to be in a state. The first and simplest is based on the value of a quality (e.g., color). Second, a TA's configuration could also be considered a state (e.g., a valve is open or closed). The third type of state results from the interaction of several largely independent components in a Microworld (e.g., a traffic jam). RQs support the description of qualities which are shared among two or more Independent

---

[6] In the strictest reading of BFO, Object Aggregates are physically connected Material Entities [24] without necessarily any functional relationships. We adopt a broader interpretation in which the Object Aggregate is a collection of Parts. In comparison, [21] describes a symphony orchestra as an example of an Object Aggregate without explaining how the relationships in the orchestra make it an Object Aggregate.
[7] This is generally consistent with Simon's notion of nearly decomposability systems [19].
[8] In [5] we introduced Thick Independent Continuants; here, we focus on Thick Aggregates.
[9] We expect to address in future papers challenges specific to the description of parts.

Continuants. Potentially, RQs could specify configurations among parts. Furthermore, a change in configuration could be implemented by a Transitional changing the RQs of the parts.

## 3 TRANSITIONALS AND MECHANISMS

### 3.1 Transitionals

A mechanism results in state changes. While BFO2 does not support state changes, we have proposed Transitionals [5, 6] which allow state changes as part of the Model-Layer. Transitionals are like Activities or Causings for ([16] p6):

> Activities are types of causes. Terms like "cause" and "interact" are abstract terms that need to be specified with a type of activity and are often so specified in typical scientific discourse.

The Transitional can be general or specific. For instance, it may describe a decomposition reaction (see Section 4.1) as breaking covalent bonds and forming ions, or it might describe specific reactions for each kind of atom. The result of the transition is clear just from knowing the initial TA and the transition, but, for clarity, we may define input-transition-output as a transitional unit. Indeed, we might develop a collection of transitional units that includes details of the circumstances under which they occur. In particular, mechanisms are chained transitions [6] whose cumulative history needs to be tracked.

We do not have to know the details of how a valve works to say that it is open. Similarly, we can say that a vehicle moves an engine without specifying whether it is a gasoline, diesel, or electric engine. When more detail is provided, we can add transitional units. For instance, if we learn that it is a gasoline engine, we can be confident that it has pistons which react to a spark, but we still may not know whether it has a carburetor or fuel injection. In effect, the transitional units can be nested. This is expressed by ([16] p13) as:

> Mechanisms occur in nested hierarchies and the descriptions of mechanisms in neurobiology and molecular biology are frequently multi-level. The levels in these hierarchies should be thought of as part-whole hierarchies with the additional restriction that lower level entities, properties, and activities are components in mechanisms that produce higher level phenomena.

Machamer, Darden, and Craver ([16] p13) also suggest that it is possible to drill down through several levels of mechanisms and then to bottom out.

> Nested hierarchical descriptions of mechanisms typically *bottom out* in lowest level mechanisms. These are the components that are accepted as relatively fundamental or taken to be unproblematic for the purposes of a given scientist, research group, or field. Bottoming out is relative: Different types of entities and activities are where a given field stops when constructing mechanisms.

It is an important insight that research communities adopt a variety of strategies to limit dealing with minimally relevant details. In any case, with richer modeling, the bottom layers for each of the fields, and thus the fields themselves, may be better coordinated.

There are many interpretations of causation. [13] observes that "it is a strength that the mechanistic account of explanation is compatible with a number of different accounts of causation". As mentioned in Section 1, in this paper, we focus on relatively straightforward mechanisms having deterministic causation and generally follow an operationalist approach [27]. On the other hand, causation for mechanisms in social science is contentious (e.g., [12, 13]) and much more work is needed before attempting to develop a descriptive framework for social mechanisms.

### 3.2 Functionality

Functionality may be described as "what something does". In BFO2 a function is a Dependent Continuant associated with an Independent Continuant ([22] p103) state:

> Functions are ontologically challenging in that they are generally taken to involve a normative element: for a thing to *have a function* is for it to have something it is supposed to do and that it can do more or less well.

However, it is also possible to talk about the function of an activity or an entire mechanism ([16] p6):

> It is common to speak of functions as properties "had by" entities, as when one says that the heart "has" the function of pumping blood or the channel "has" the function of gating the flow of sodium. This way of speaking reinforces the substantiality tendency against which we have been arguing. Functions, rather, should be understood in terms of the activities by virtue of which entities contribute to the workings of a mechanism. It is more appropriate to say that the function of the heart is to pump blood and thereby deliver (with the aid of the rest of the circulatory system) oxygen and nutrients to the rest of the body. Likewise, a function of sodium channels is to gate sodium current in the production of action potentials. To the extent that the activity of a mechanism as a whole contributes to something in a context that is taken to be antecedently important, vital, or otherwise significant, that activity too can be thought of as the function of the mechanism as a whole.

We prefer associating functions with Transitionals rather than Independent Continuants, but both are possible. A difficulty for BFO2's definition of function is that there is no simple structured way to specify what the Independent Continuant does. By comparison, with our approach to modeling, a mechanism can be used to show what the function does. While functions can be judged normatively, we propose that normativity should be measured by the number of other mechanisms and systems in which they participate in a given scenario.

Finally, it is useful to distinguish among different three types of mechanisms and their functions. First, some mechanisms are systematically designed to perform a specific function. Mechanical engineers and system designers have developed a number of structured approaches (such as FBS and SADT) for developing specific functionality in their designs. The second type of mechanism evolves to fill a niche. Thus, biological mechanisms often have clear functionality, but may involve the interaction of apparently irrelevant factors. Understanding these complexities is a goal of biological research [8, 9].[10] The third type of mechanism is best seen as an aspect of a natural physical system. For example, we can identify the mechanism of a geyser and argue that the function of a geyser is to relieve water pressure built up inside the earth.

## 3.3 Systems

A system is "a complex object whose parts or components are held together by bonds of some kind" ([10] p188). The notion of a system is also one of the foundations of computing [19, 20]. General Systems Theory [24] describes a number of dimensions for different types of systems. We consider open and closed systems in the following section. It is also useful to distinguish between systems with local control as compared to those with centralized control (or, possibly, both). Some systems, like an automobile, serve a specific function and clearly incorporate multiple mechanisms. Other systems, like the Solar System, have no obvious input or output, or function.

## 3.4 Microworlds

The term Microworld is adopted from object-oriented modeling. There are two distinct functions for Microworlds. The first is as a framework or scope for allowing the interaction of autonomous interacting agents. A Microworld can also support implementation details such as a scheduler for determining the order in which the status of the TAs is evaluated. For example, in modeling city traffic, the model could include independent TAs such as traffic lights and cars. Some of the changes in the TAs are internal (e.g., the changing color of the traffic lights (cf., [5])) while other interactions are with independent TAs (e.g., cars stop when the traffic light turns red). This fits well with the expectation of encapsulation in object-oriented systems. The interaction between the independent TAs can be implemented as message passing.

It should be possible to have Microworlds without spatial constraints, but Microworlds will often include spatial relationships. The Microworld concept allows us to specify axioms such as gravitational attraction. A

---

[10] Designing mechanisms often follow a rigorous methodology. The design process may apply functional programming or the SOLID principles of Object-Oriented Modeling. While naturally evolved systems tend to be modular [18], they do generally do not reach the standards of functional programming.

Microworld may be used to develop structured scenarios to describe several interacting mechanisms such as the mechanisms in a cell or procedures associated with a person's employment.[11]

Systems and Microworlds overlap. A system may occupy a Microworld. For instance, a cell could be considered as a System that could be implemented in a Microworld. Presumably, a Microworld as a whole is a closed system. However, it may be possible to specify some subsets of a Microworld as open systems.

## 4 EXAMPLES

### 4.1 $2H_2+O_2 \rightarrow 2H_2O$

#### 4.1.1 Deterministic Model

Figure 1 is a schematic of a common phenomenon: the formation of water from hydrogen and oxygen molecules.

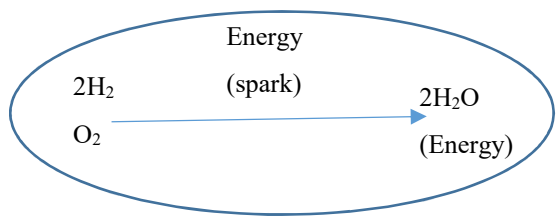

Figure 1. The phenomenon: A synthesis reaction yielding water.

Figure 1 shows the phenomenon, but not a full mechanism. Figure 2 extends Figure 1 as two transitions. The first is the decomposition of $H_2$ and $O_2$ molecules into $H^+$ and $O^{--}$ ions. The second combines those ions into $H_2O$ molecules. Certain conditions must be met for the transition to occur. The spark increases the energy of the $H_2$ and $O_2$ molecules. That energy serves as a Transitional and causes the disassociation of the molecules into ions. The ions then combine to form the $H_2O$ molecules, and energy is released.

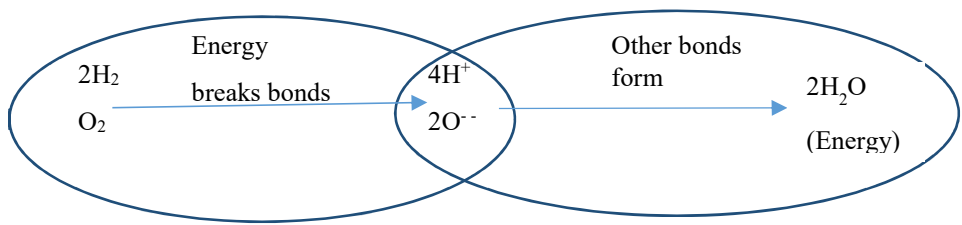

Figure 2. The mechanism in the reaction in Figure 1 can be divided into a decomposition reaction and a combination reaction. It is essential that the result of the first transition match the input required for the second transition.

However, there are challenges in developing a model of the phenomenon consistent with BFO2. There are molecules, atoms, and ions involved. A molecule Aggregate is part of a water Aggregate. Atoms are also Aggregates; they are composed of electrons, protons, and neutrons. On the other hand, this phenomenon could be modeled in an XFO Microworld. In fact, it is implicit that the energy from the spark must act concurrently on several independent entities (the $H_2$ and the $O_2$ molecules).

#### 4.1.2 Stochastic Models

As satisfying as Figure 2 is, it leaves out a great deal. The mechanism is idealized because it does not show the (relatively low) possibility of the ions recombining as $H_2$ and $O_2$ molecules. Nor does it include constraints such as the density of the gases. Moreover, in the real world, such reactions occur in the context of a very large number of other molecules. A related example is the phase changes of water. When a pot of water is being heated, the heat disrupts hydrogen bonds within the water molecules. But the state change of the water coming to a boil is observed for the entire pot of water. It is impossible to model all the discrete interactions in many cases. Thus models may require some generalization operator. In addition, it should be helpful to use multiple models [25] to describe different aspects of mechanisms.

---

[11] In addition to mechanisms, there are other closely related structured, sequential activities such as procedures and workflows that are similar to mechanisms although they may involve some level of human intervention.

## 4.2 Water Tank Drain and Refill

As another example, we consider draining and refilling a water tank.[12] This introduces another set of modeling issues. Once the tank has been drained, it needs to be reset to its initial state before it can be drained again. This could be considered as two mechanisms (drain and refill). It makes sense to describe the two interrelated mechanisms in one model, but that also means that the model needs to allow concurrency.

Figure 3 illustrates the tank mechanism using a Petri Net. Petri Nets are varied, flexible and widely used. Transitions in Petri Nets are controlled by tokens; when all the necessary conditions are met as indicated by the accumulation of a complete set of tokens, the transition is triggered and new token fired.

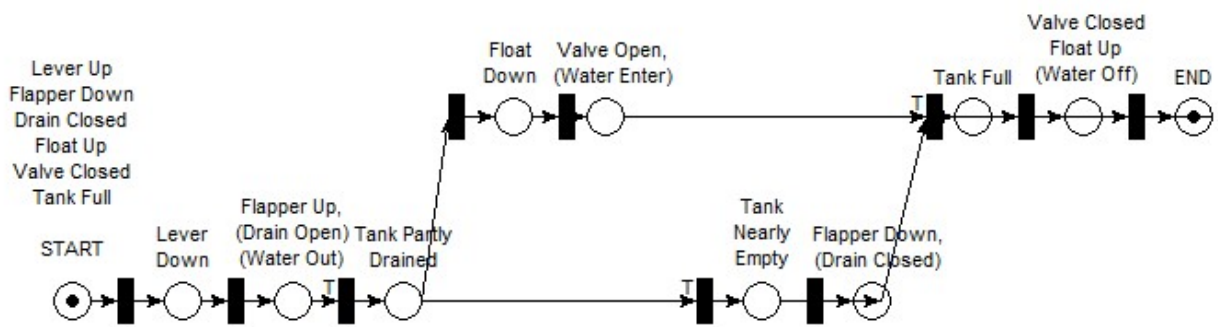

Figure 3. A Petri Net is showing a chain of transitions for the water tank drain and refill. There are two symmetric mechanisms, which are coordinated. Some of the transitions are concurrent; a full model would include a timing parameter. The time delay, T, is related to the water pressure.

While the Petri Net shows the flow-through states nicely, a number of aspects needed for a full representation of mechanisms are not included. For instance, we do not know the kinds of transitions that occur. Although they have a straightforward notation, Petri Nets are not as expressive as a programming language. Ultimately, the tank mechanism model should be implemented into XFO, compiled, and all transitions validated.

## 5 SEMANTIC MODELING FRAMEWORK (SMF)

While it should be possible to implement mechanisms with only the core components (see Section 2.2), it will also be helpful to document the mechanisms more fully than only with core components. Here we outline a more complete descriptive framework for a machine-processable framework for describing mechanisms. For example, a knowledgebase [4] of TAs could include different types of atoms and ions, and a knowledgebase of transitional units could include mechanisms associated with different types of vehicle engines. Ideally, these descriptions would be in some structured format, but text may often be more practical. Sections 5.1, 5.2, and 5.3 are collections of proposed descriptions.

## 5.1 Mechanism Metadata

**Type of Mechanism:** Simple Linear, Cyclic, Concurrent, Feedback, Continuous, Stochastic, Asynchronous. In this paper, we focus on simple linear and concurrent mechanisms.

---

[12] We might call this a procedure because it starts with a human action, though, in this case, the difference seems minimal.

**Type of Model:** As described above, representations of mechanisms are necessarily simplified. The nature of this simplification should be described. There should also be indicators of the completeness and simplicity of models [25].

**Function Type:** Designed, Evolved, Natural, None Apparent.

**Dynamic Elements in Execution:** Because the mechanism is a sequence of transitions, transitions early in the sequence can affect those that come later. These sequential constraints should be noted.

**Context:** Ideally, the description of a mechanism would include a full description of the conditions in which it operates. This could include generic descriptions (e.g., in vivo), immediate causes, or common assumptions (e.g., chemical measurements are often made at standard temperature and pressure, STP). A full representation would specify the complete set of necessary conditions for each transition. In addition, some mechanisms may occur under highly unusual circumstances (e.g., in black holes) or in other eras (e.g., ancient Egyptian embalming techniques).

**Base Metadata:** Author, Date, Version.

## 5.2 Core Components

**Phenomenon:** The phenomenon is what the mechanism does. Figure 1 is an example; it describes the transition while a more complete view of the overall mechanism is shown in Figure 2.

**Parts:** Parts should be modular and functional parts should have a specific role in the mechanism (see Section 2.1).

**Causings and Organization:** The organization is the sequence of states of the mechanism. We interpret causation as the transitions used for scientific descriptions.

## 5.3 Additional Elements

**Explanations:** While the types of descriptions in the previous section should allow implementation of the mechanism, they do not necessarily enable a clear account for why something happened. This field allows for such explanations.

**Variations, Analogs, and Contrasts:** A given mechanism may have alternatives or analogs (e.g., in related species, across cultures, etc.). This field could also include a statement about possible reasons for these differences.

**Implications/Applications:** Mechanisms that have a substantial impact are particularly noteworthy. However, in some cases developing a structured representation of those implications will be challenging because of the breadth of the possible scenarios. The description could also include a description of what happens if the mechanism is broken.

**Evidence:** The descriptions of mechanisms could be linked to the structured evidence for why we believe particular factors are involved (cf., [1, 17]).

# 6 CLOSING

Mechanistic theories in science have been widely considered over the past 30 years. We attempt to draw a modeling framework from that discussion. Specifically, we have outlined a Semantic Modeling Framework (SMF) for structured descriptions of mechanisms. One application we envision is direct representation knowledgebases of scientific research reports [1, 4]. In addition to the representation of mechanisms, these knowledgebases would include structured research workflows, results, and analysis. For instance, as suggested in [1], a knowledge-based-report might include a test of the research results such as:

```
if ((NAD = =LOW) && (Degeneration = = LOW)) then {prefer ConceptualModel1;}
else if ((NAD = =LOW) && (Degeneration = =NORMAL)) then {prefer ConceptualModel2;}
```

As noted in Section 1, this paper focuses on relatively simple mechanisms. There are many challenges in describing more complex mechanisms. For instance, the beating of a heart might be described as a cyclic mechanism with multiple, coordinated transitionals. Moreover, while simple procedures should be relatively easy to model in a scenario, social mechanisms are much more challenging [13]. The example of a symphony orchestra as an Object Aggregate is especially challenging given that it involves considerable coordination among a group of people over an extended time. While aspects of this approach still need to be refined, the structured framework presented here brings us closer to the broad goal of model-oriented information organization [2, 3].